\documentclass[reprint,amsmath,amssymb,aps,prl,superscriptaddress]{revtex4-2}
\usepackage{graphicx}
\usepackage{dcolumn}
\usepackage{bm}
\usepackage[colorlinks,linkcolor=blue,urlcolor=blue,anchorcolor=black,citecolor=blue]{hyperref}
\usepackage{url}
\usepackage{subfigure}

\begin{document}
	
\title{Self-supervised graph neural networks for accurate prediction of N\'{e}el temperature}

\author{Jian-Gang Kong}
\affiliation{School of Science, Chongqing University of Posts and Telecommunications, Chongqing 400065, China}\author{Qing-Xu Li}
\affiliation{School of Science, Chongqing University of Posts and Telecommunications, Chongqing 400065, China}%
\affiliation{Institute for Advanced Sciences, Chongqing University of Posts and Telecommunications, Chongqing 400065, China}%
\author{Jian Li}
\affiliation{School of Science, Chongqing University of Posts and Telecommunications, Chongqing 400065, China}%
\affiliation{Institute for Advanced Sciences, Chongqing University of Posts and Telecommunications, Chongqing 400065, China}
\affiliation{Southwest Center for Theoretical Physics, Chongqing University, Chongqing 401331, China}
\author{Yu Liu}
\affiliation{Inspur Electronic Information Industry Co., Ltd, Beijing 100085, China}
\author{Jia-Ji Zhu}
\email{zhujj@cqupt.edu.cn}
\affiliation{School of Science, Chongqing University of Posts and Telecommunications, Chongqing 400065, China}%
\affiliation{Institute for Advanced Sciences, Chongqing University of Posts and Telecommunications, Chongqing 400065, China}
\affiliation{Southwest Center for Theoretical Physics, Chongqing University, Chongqing 401331, China}



\begin{abstract}
Antiferromagnetic materials are exciting quantum materials with rich physics and great potential for applications. It is highly demanded of the accurate and efficient theoretical method for determining the critical transition temperatures, N\'{e}el temperatures, of antiferromagnetic materials. The powerful graph neural networks (GNN) that succeed in predicting material properties lose their advantage in predicting magnetic properties due to the small dataset of magnetic materials, while conventional machine learning models heavily depend on the quality of material descriptors. We propose a new strategy to extract high-level material representations by utilizing self-supervised training of GNN on large-scale unlabeled datasets. According to the dimensional reduction analysis, we find that the learned knowledge about elements and magnetism transfers to the generated atomic vector representations. Compared with popular manually constructed descriptors and crystal graph convolutional neural networks, self-supervised material representations can help us obtain a more accurate and efficient model for N\'{e}el temperatures, and the trained model can successfully predict high N\'{e}el temperature antiferromagnetic materials. Our self-supervised GNN may serve as a universal pre-training framework for various material properties.
\end{abstract}

\maketitle

\emph{Introduction.---}Antiferromagnetic material is an exciting class of quantum materials with rich physics in condensed matter theory and many practical applications. In theoretical aspects, antiferromagnetic materials are the parent materials of a series of physical phenomena, such as high-temperature superconductivity\cite{Nagaosa}, spin liquids\cite{Kanoda}, quantum anomalous Hall effect\cite{Qiao}, and topological axion insulator\cite{LiuC}; In the application aspects, the antiferromagnetic materials can be used for implementing spin valves\cite{ParkB}, colossal magnetoresistive effect\cite{QiuZ}, room-temperature electrical switching\cite{WadleyP}, fast magnetic moment dynamics\cite{JungwirthT}, and other antiferromagnetic spintronic applications\cite{BaltzV}. The most crucial parameter for antiferromagnetic materials is the N\'{e}el temperature, which marks the antiferromagnetic ordering transition, similar to the Curie temperature of a ferromagnet. The experimental determination of the N\'{e}el temperature often requires long periods and high costs, while the theoretical predictions with analytical or numerical approaches are highly non-trivial. It is usually quite tedious or challenging to specify the microscopic Hamiltonian corresponding to the antiferromagnetic material and determine the type and strength of magnetic interactions\cite{LiX}. The most ordinary theoretical methods are based on mean-field theory, which is not very effective due to the divergent length scale near the critical point. The quantum Monte Carlo method suffers the well-known negative sign problem\cite{LohE}, and the powerful tensor network method bears difficulty in dealing with physical properties at finite temperatures in two and higher dimensions\cite{DidierP,Gong2021,CincioL}. Therefore, it is of great significance to develop methods that can predict the N\'{e}el temperature both accurately and efficiently. 

Machine learning aims to fit a predictive model or to find patterns from data, which has already achieved great success in predicting physical quantities\cite{SchmidtJ} and inverse-designing materials\cite{NohJ}. Recently, graph-based neural networks\cite{XieT,ChenC,SaucedaH} demonstrate the state of the art prediction performance in various material properties with a large amount of data. Especially, crystal graph convolutional neural network (CGCNN)\cite{XieT} is used to fit a function between the graph representation of materials and the target material properties. Based on the flexible CGCNN framework, one can either encode more physical information into the crystal graph\cite{KaramadM,ParkC} or further optimize the edges\cite{ParkC}.

In predicting the transition temperature of magnetic materials with machine learning methods, most studies focus only on the Curie temperature of ferromagnetic materials. For example, Nelson et al. showed the model of best performance among the random forest, kernel ridge regression (KRR), and neural network model achieves a mean absolute error (MAE), 57 K, on a ferromagnetic material dataset of size 2,500\cite{NelsonJ}. The random forest model trained by Long et al. reaches an accuracy of 0.87 to distinguish 1,749 ferromagnetic and 1,056 antiferromagnetic materials, while the regression analysis of Curie temperature shows that the MAE is about 55 K\cite{LongT}. Other efforts focus on the impact of different inputs on the prediction accuracy, for instance, constructing 21 descriptive variables as inputs of the KRR algorithm to predict the Curie temperature of transition metal-rare earth compounds\cite{NguyenD}. However, there are very few reports about predicting N\'{e}el temperature by machine learning, such as a support vector regression (SVR) model trained on perovskite manganese oxides of size 127, which achieves an root mean squared error of 32.3K on a test set of size 32\cite{LuK}.

The scarcity of data is the grave difficulty in machine learning for the transition temperature of magnetic materials. There is still no large-scale computational dataset for magnetic materials with calculated transition temperatures. A recent work\cite{CourtC} has mined a large number of published papers for data and constructed a magnetic phase transition temperature dataset of size 40,000, utilizing tools from the field of natural language processing and relation extraction techniques. Unfortunately, the complete material structures are absent in the constructed dataset, where the labels also lack quality assurance. Due to the scarcity of magnetic material data, most of the machine learning studies on magnetic materials make use of the ensemble learning algorithms, such as random forests, rather than deep learning algorithms, such as graph neural networks (GNN), on small dataset\cite{DunnA}. Yet the performance of the former highly relies on the quality of material descriptors\cite{GhiringhelliLM}.

Self-supervised learning is a new method to train neural networks without the need for expensive labels, utilizing supervised signals from the content or the intrinsic structure of the data itself. As a pre-training strategy for models with many parameters, self-supervised learning has gained significant interests in natural language processing\cite{DevlinJ} and computer vision\cite{HeK}. In the graph machine learning field, it is proposed that a pre-training method on large datasets can recover the randomly masked elemental information in molecular graphs and improve prediction performance on various small molecular datasets\cite{HuW}. Since the crystal graphs representing crystalline materials contain rich physical information, it is believed that different types of self-supervised learning on large-scale crystal graphs can capture prior solid knowledge and lead to performance improvement in downstream tasks.

In this Letter, we propose to combine the representation learning capabilities of GNN with the efficiency of the standard machine learning model on the N\'{e}el temperature dataset. We extract high-level representations of materials in a self-supervised manner by training the GNN to reproduce elemental information and magnetic moments. As the input of the regression model, self-supervised material descriptors can outperform popular material descriptors and the powerful CGCNN, due to their simplicity, high-relevance, and low computational cost. The trained KRR model possesses the capability to screen magnetic materials with high N\'{e}el temperatures from database for spintronic applications.

\emph{Self-supervised learning on crystal graphs.---} The crystal graph is a multi-graph characterized by node vectors, edge vectors, and adjacent matrix, allowing multiple edges between the same pair of nodes due to the periodicity of crystalline materials. The nodes of the graph encode elemental information of atoms, such as the period number and the group number. The edges of the graph encode the distance between atoms which implicitly capture interactions or bondings. The node vector $\mathbf{v}_{i}$ in the crystal graph is updated by the $t$-th GNN layer as follows,
$$
\mathbf{v}_{i}^{(t+1)} = \mathbf{v}_{i}^{(t)} + \sum_{j,k}\sigma\left(\mathbf{z}_{(i,j)_{k}}^{(t)}\mathbf{W}_{f}^{(t)}+\mathbf{b}_{f}^{(t)}\right)\odot g\left(\mathbf{z}_{(i,j)_{k}}^{(t)}\mathbf{W}_{s}^{(t)}+\mathbf{b}_{s}^{(t)}\right).
$$
The node $\mathbf{v}_{i}$ updates itself by aggregating the messages provided by its neighbors, which belongs to the message passing mechanism\cite{GilmerJ}. $\mathbf{W}_{f}$ and $\mathbf{W}_{s}$ are learnable weight matrices acting on a pair of neighbors $\mathbf{z}_{(i,j)_{k}}=\mathbf{v}_{i}\oplus\mathbf{u}_{(i,j)_{k}}\oplus\mathbf{v}_{j}$, where $\mathbf{u}_{(i,j)_{k}}$ represents the vector corresponding to the $k$-th edge between node $i$ and node $j$, and $\oplus$ denotes the concatenation of two vectors. $\mathbf{b}_{f}$ and $\mathbf{b}_{s}$ are the corresponding biases of linear transformations. The $\sigma$ and $g$ are non-linear activation functions that can increase the expressive power of neural networks and filter out the most important bonds during training. The $\odot$ denotes the element-wise product, or the Hadamard product, between two vectors.

\begin{figure}[htbp]
\centering
\subfigure[]{
	\begin{minipage}[t]{1.0\linewidth}
		\centering
		\includegraphics[height=4.0cm,width=9.0cm]{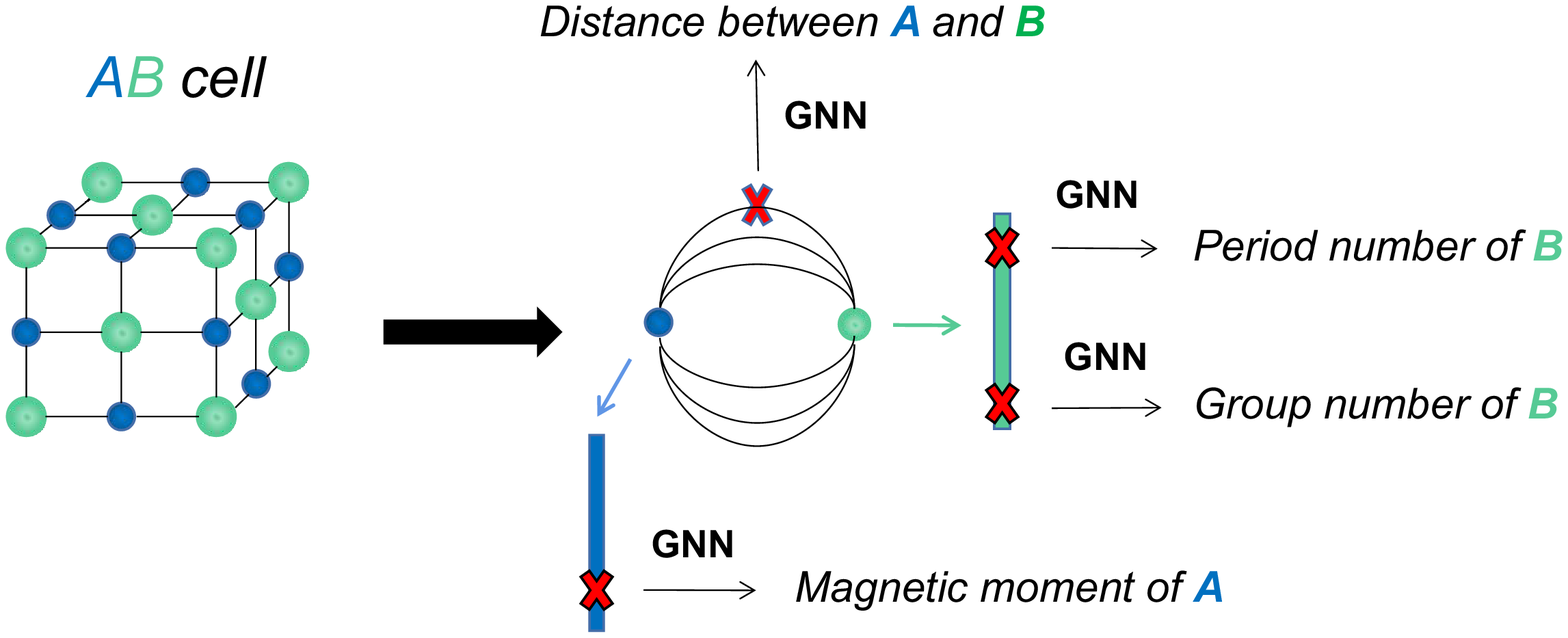}
	\end{minipage}
	\label{SL1}
}
\hspace{.10in}
\subfigure[]{
	\begin{minipage}[t]{1.0\linewidth}
		\centering
		\includegraphics[height=4.0cm,width=9.0cm]{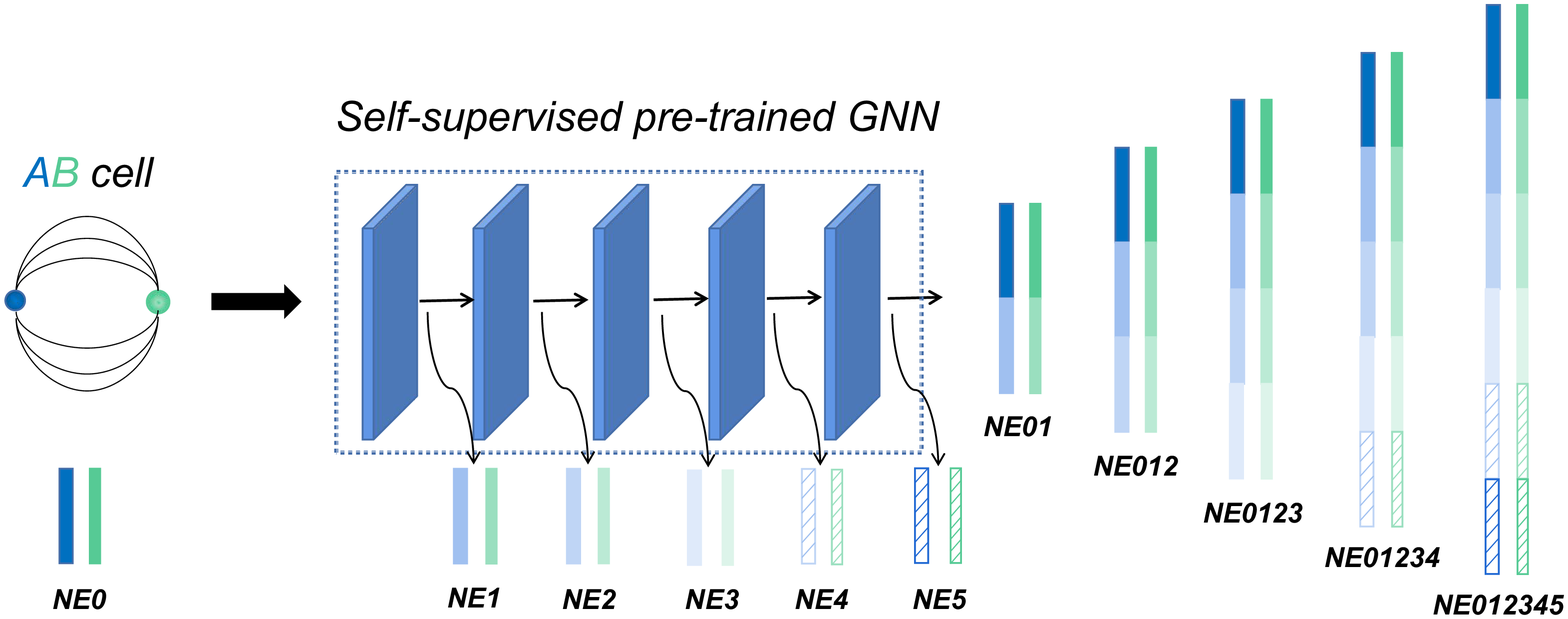}
	\end{minipage}
	\label{GE1}	
}
\caption{(a) Self-supervised training on crystal graphs (AB cell as an illustrative example). The green (blue) balls represent B (A) atoms, the green (blue) bar is the node vector corresponding to B (A) atom, and the red crosses denote randomly masked node or edge information during training. A GNN is trained to recover the masked information based on the surrounding crystal environment. (b) 64-dimensional atomic vectors generated from the self-supervised pre-trained GNN, given a crystal graph as input (AB cell as an illustrative example). The green (blue) bars at the bottom are single-scale atomic vectors of different layers, while those on the right are multi-scale atomic vectors.}
\end{figure}

In order to make full use of the sizeable unlabeled dataset merely with information about elements and structures, we propose to extract chemical rules by self-supervised learning on a computational material dataset of size 60,000, constructed from the Materials Project\cite{JainA}. More specifically, as shown in Fig.\ref{SL1}, we randomly mask the elemental information, the group number and the period number, of a pre-defined proportion of atoms in the unit cell during training. A GNN is then trained to recover the information based on the surrounding crystal environments of the masked atoms. Also, we randomly mask the edge information, the distance between atoms, of the masked nodes, and train the neural network to correctly predict the distance information. We expect GNN to learn high-level knowledge of crystal structures and chemical properties by self-supervised training and save it in the form of neural network weights. The trained GNN is denoted as Elem-GNN.

We also perform self-supervised learning on a computational magnetic material dataset of size 50,000, constructed from the Materials Project, to extract knowledge about magnetism. Similar to the self-training process described above, we still randomly mask the elemental and distance information of each material. Some of the masked atoms have non-zero magnetic moments in this case. Therefore, we can train another GNN, which we refer to as Mag-GNN, to reproduce the size of the magnetic moments of masked atoms. 

Given the original crystal graphs as inputs, we can generate 64-dimensional node embeddings (NE) by utilizing the self-supervised pre-trained GNN, i.e., Elem-GNN and Mag-GNN, as shown in Fig.\ref{GE1}. We name the atomic vectors Elem-NEs and Mag-NEs from the Elem-GNN and Mag-GNN respectively. We can transfer the knowledge obtained by self-supervised training from GNN to the generated atomic representations. By averaging over the NEs in the same unit cell, we can obtain the vector representation, or the graph embeddings (GE), of the materials, which can be directly taken as the input vectors of machine learning models for studying material properties. We use a five-layer GNN in the self-supervised training and output the corresponding self-supervised atomic vector NE$i$, ($i=0,1,2,3,4,5$) from each layer. Then we obtain multi-scale atomic representations NE01, NE012, ..., and NE012345 by concatenating atomic representations of different layers and thus obtain multi-scale material representations GE01, GE012, ..., and GE012345. The necessities of the multi-scale representations are two-fold: (1) The GNN updates the atomic representations through the message passing mechanism, and the atomic vectors of the deeper GNN layer receive a more extensive range of environmental information; (2) The deep GNN, however, suffers the so-called over-smoothing problem\cite{LiQ}, which results in the degradation of the performance from the similarities of the node representations of deep GNN layer. And the CGCNN is also perplexed by the same problem due to the sizeable effective range of single message passing for unit cells with a few atoms. Multi-scale material representations can reduce the similarities between the atomic representations and offer us additional freedom by controlling the amount of environmental information in descriptors.

\emph{Dimensional reduction visualization.---} To examine the effects of self-supervised training and to make sure that the learned knowledge is transferred to atomic vectors, we visualize the distribution of two-dimensional (2D) points corresponding to the 64-dimensional self-supervised atomic vectors Elem-NEs, Mag-NEs, and randomly initialized atomic vectors Random-NEs by utilizing the t-SNE (t-distributed stochastic neighbor embedding) technique\cite{vanL}. The 2D data points obtained by dimensional reduction of Elem-NEs are labeled with element types, and those of Mag-NEs are marked with the size of magnetic moments. We expect that the patterns formed by the self-supervised atomic vectors are more regular than the random atomic vectors, which highlights the benefits of self-supervised learning. The magnetic moment dataset used for visualization is constructed from MAGNDATA\cite{GallegoS,ElcoroL}, containing 1,816 magnetic atoms from 29 elements, including 13 transition metals and 12 lanthanides. 

\begin{figure}[htbp]
	\centering
	\subfigure[]{
		\begin{minipage}[t]{0.4\linewidth}
			\centering
			\includegraphics[height=4.0cm,width=4.0cm]{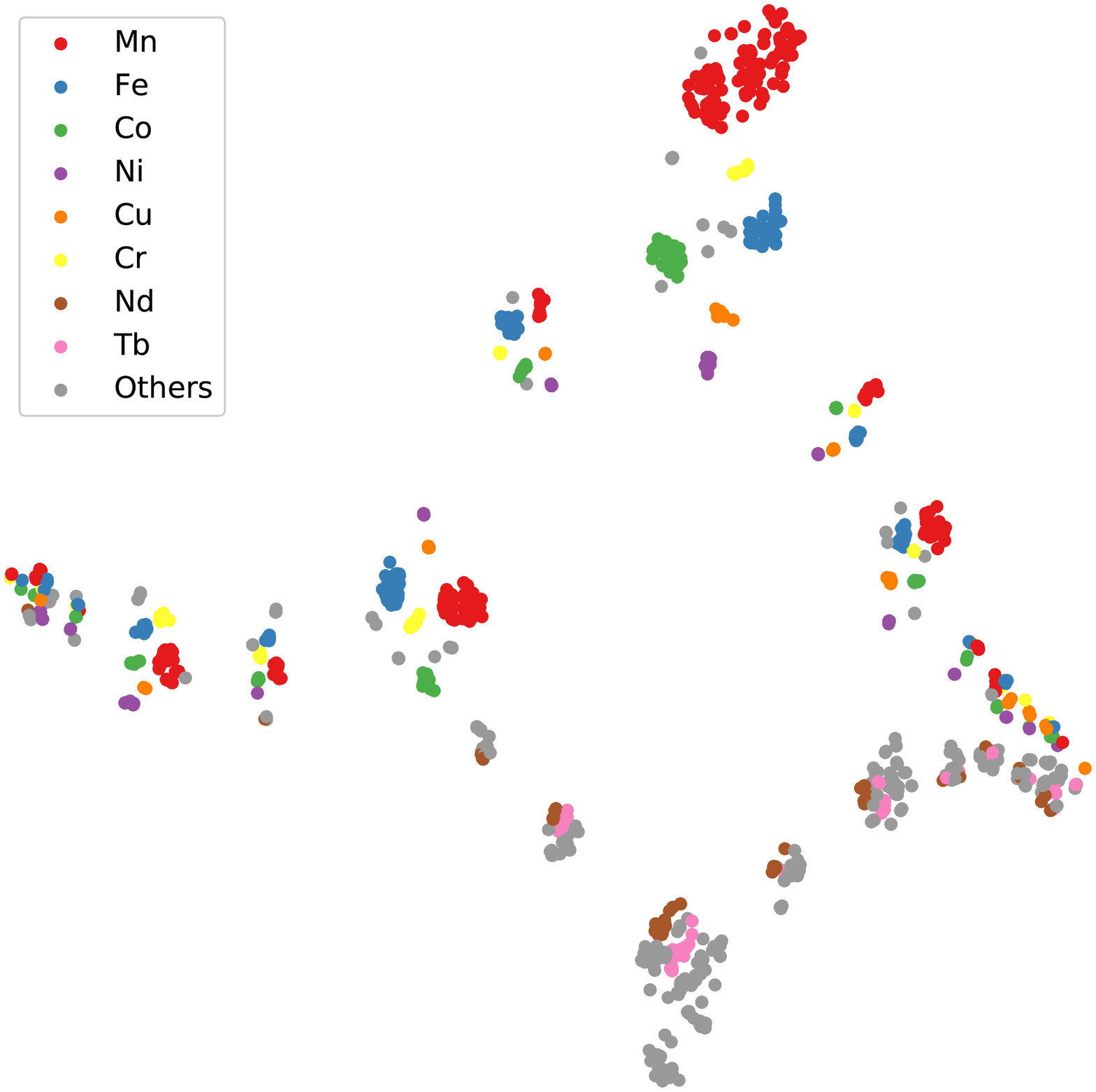}
			\label{RE}
	\end{minipage} }
	\hspace{.20in}
	\subfigure[]{
		\begin{minipage}[t]{0.4\linewidth}
			\centering
			\includegraphics[height=4.0cm,width=4.0cm]{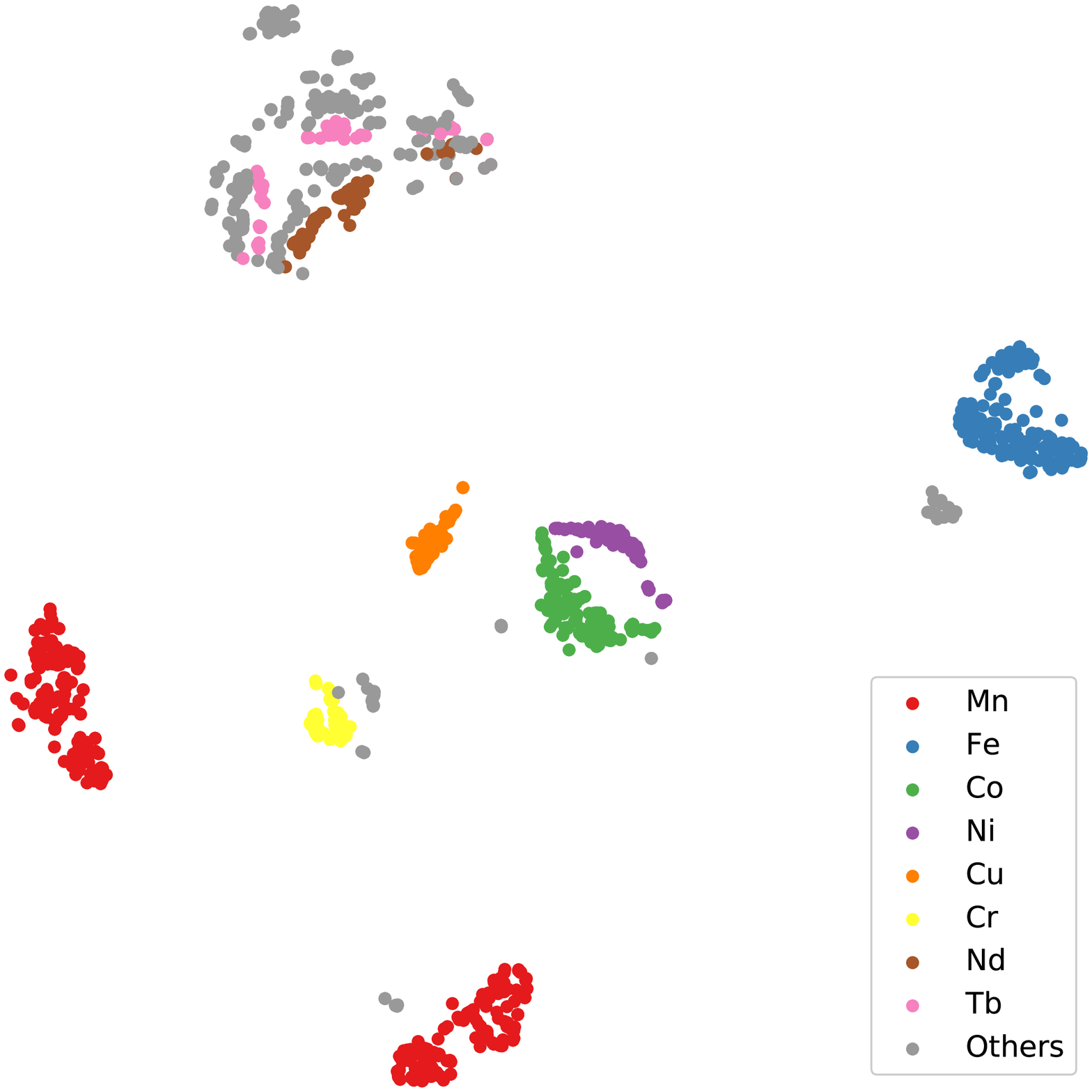}
			\label{SE}
	\end{minipage}}
	\subfigure[]{
		\begin{minipage}[t]{0.4\linewidth}
			\centering
			\includegraphics[height=4.0cm,width=4.0cm]{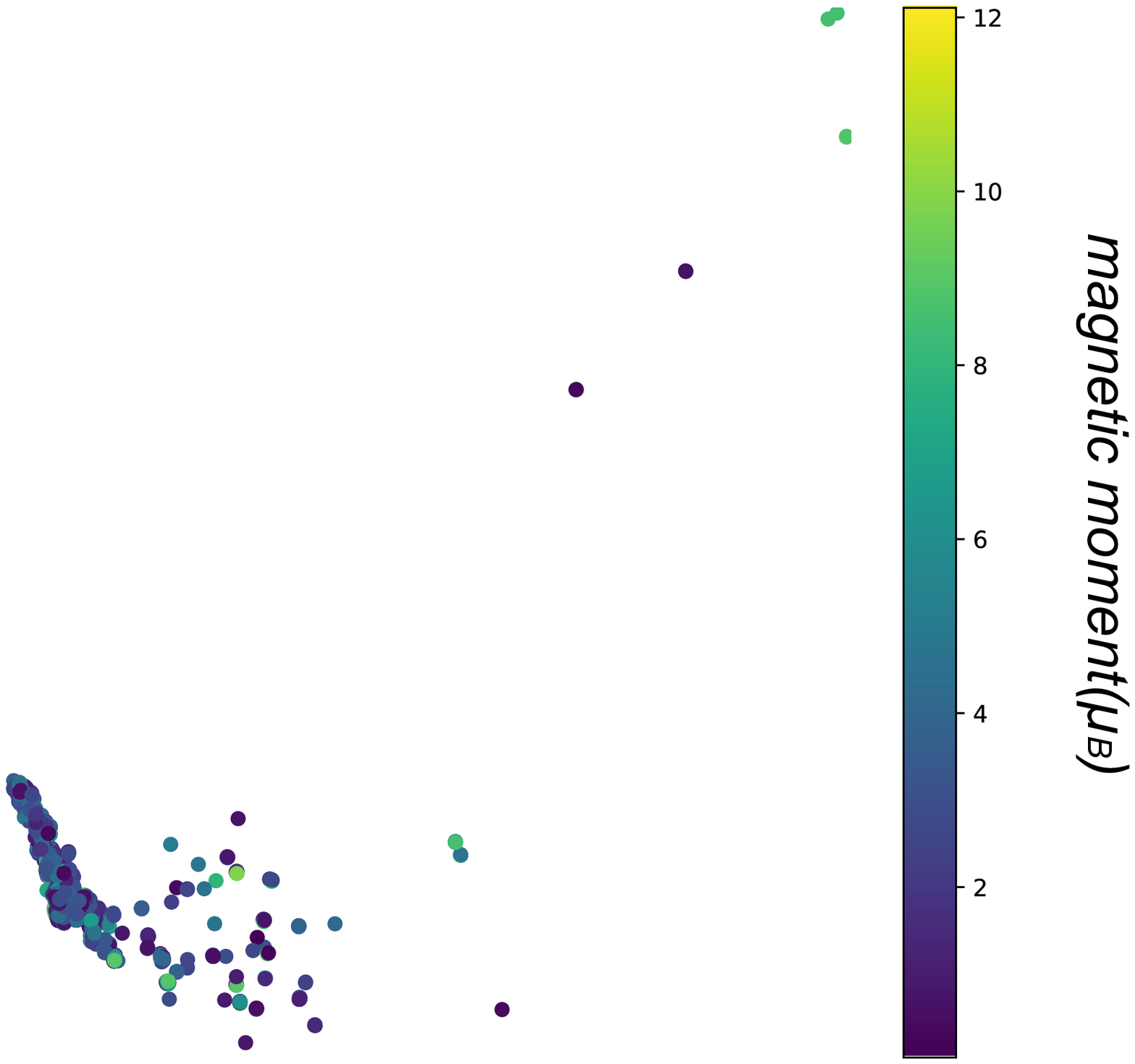}
			\label{RM}
	\end{minipage} }
	\hspace{.20in}
	\subfigure[]{
		\begin{minipage}[t]{0.4\linewidth}
			\centering
			\includegraphics[height=4.0cm,width=4.0cm]{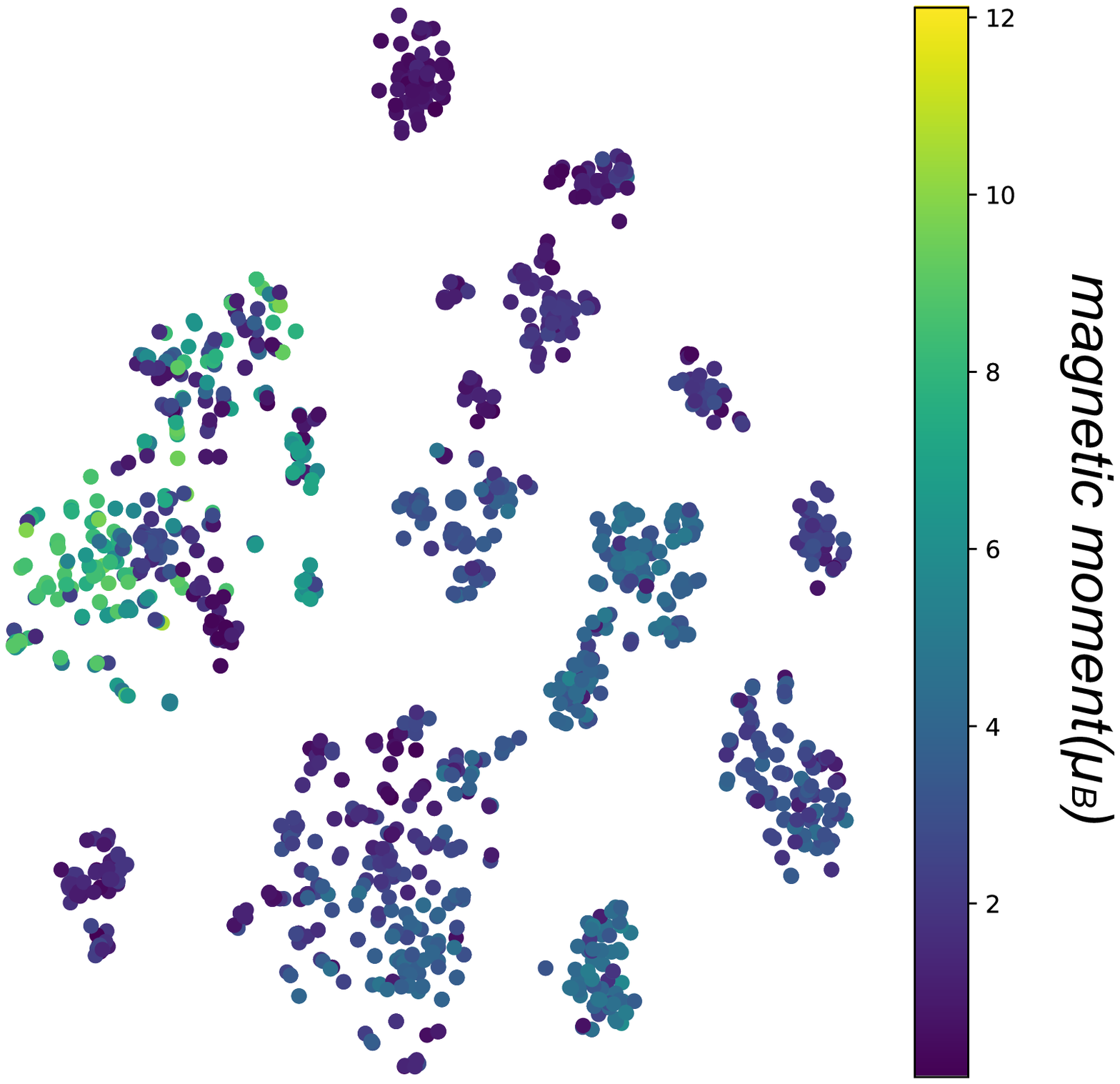}
			\label{SM}
	\end{minipage} }
	\caption{t-SNE visualization of self-supervised atomic vectors Elem-NEs, Mag-NEs, and Random-NEs labeled by element types and magnetic moments. Different colors represent different elements or moments.(a) 2D distribution of Random-NEs under element labels. (b) 2D distribution of Elem-NEs under element labels. (c) 2D distribution of Random-NEs labeled by magnetic moments. (d) 2D distribution of Mag-NEs labeled by magnetic moments.}
\end{figure}

From Fig.\ref{RE} and Fig.\ref{SE}, we can see that the atomic vectors Elem-NEs form more well-clustered patterns under element labels than Random-NEs, especially for transition metals. The most abundant Mn atoms form several smaller clusters, which may result from competition between different local crystal environments. Most lanthanides, such as Nd and Tb, are also well clustered in the upper region, as shown in Fig.\ref{SE}. In contrast, Random-NEs shown in Fig.\ref{RE} have no clear organizing patterns, which indicates that Elem-NEs indeed contain richer chemical information. The distribution of magnetic moments is shown in Fig.\ref{RM} and Fig.\ref{SM}: The points of Random-NEs shrink into a small region, and the magnetic moments of different sizes are mixed; The points of Mag-NEs are more uniformly distributed, and the magnetic moments of different sizes are distinguished. For example, the magnetic moments in the range $6 \mu_{B} \sim 12 \mu_{B}$ are primarily distributed in the left middle region, while the magnetic moments in the range $4 \mu_{B} \sim 6 \mu_{B}$ are distributed in the right and bottom region. The dimensional reduction analysis verifies that Mag-NE contain more knowledge about magnetism, and we can indeed extract useful physical information from the self-supervised pre-training process.

\begin{figure}[htbp]
	\centering
	\includegraphics[height=5.0cm,width=7.0cm]{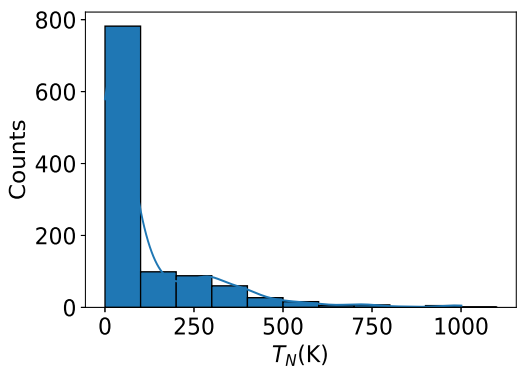}
	\caption{Distribution of experimental antiferromagnetic materials at different N\'{e}el temperature ranges.}
	\label{Dist}
\end{figure}

\emph{Prediction performance on experimental N\'{e}el temperature dataset.---} To further understand the performance of self-supervised material representations, GEs, we train and evaluate a standard machine learning model with different material representations on the experimental N\'{e}el temperature dataset. The dataset contains a total of 1,007 antiferromagnetic materials in the MAGNDATA database with corresponding magnetic structures and experimentally measured N\'{e}el temperatures. However, we can find that the distribution of materials with different N\'{e}el temperatures is very unbalanced, shown in Fig.\ref{Dist}. Most materials are in the low-temperature zone, and 881 materials are below 200 K, accounting for 80\% of the total materials. This imbalance of distribution poses tough challenges to machine learning algorithms. 

The machine learning model we employed is KRR, a nonlinear regression algorithm that combines kernel trick and ridge regression, which is widely used in material science. We compare the popular material representations, such as sine matrix (SM), Ewald sum matrix (ESM)\cite{FaberF}, and orbital field matrix (OFM)\cite{PhamT}, with our proposed multi-scale self-supervised material representations Elem-GEs and Mag-GEs. The SM and ESM can capture Coulomb interactions of crystalline materials, and their lengths are square of the maximum number of atoms in the unit cell. The OFM can encode orbital interactions of materials, and its length is 1,056, independent of the dataset. In contrast, the length of self-supervised material representations is only 64 per scale, independent of the dataset as well. Therefore, we construct a dataset of 748 materials keeping magnetic materials of a unit cell with fewer than 60 atoms for comparison, where SM and ESM are 3,600-dimensional vectors for this sub-dataset. In order to evaluate the performance of the model, we divide the dataset into a training set and a test set by 8:2 and then perform 10-fold cross-validation and grid search on the training set for the best hyperparameters. Given the optimal hyperparameters, we re-train the model on the whole training set and find the performance scores on the test set. The above procedure repeats five times with random train-test splits of dataset. The final model performance is given by the mean and standard deviation of scores on five different test sets, and the evaluation process is more reliable for a small dataset. 

\begin{table}[htbp]
	\caption{The prediction performance of the KRR model on the N\'{e}el temperature sub-dataset when the manually constructed descriptors ESM, SM, and OFM are taken as inputs}
	\centering
	\def\temptablewidth{0.4\textwidth}
	{\rule{\temptablewidth}{1.0pt}}  
	\begin{tabular*}{\temptablewidth}{@{\extracolsep{\fill}}ccccc}
		\hline
		&&ESM&SM&OFM\\
		\hline
		&MAE&95.05$\pm$10.20&86.53$\pm$4.97&75.83$\pm$6.56\\
		\hline
		&R2&0.13$\pm$0.09&0.21$\pm$0.11&0.19$\pm$0.15\\
		\hline
	\end{tabular*}
	{\rule{\temptablewidth}{0.6pt}}
	\label{Human}
\end{table}

From the results shown in Table \ref{Human}, we can see that the traditional, manually constructed material descriptors ESM and SM have poor prediction performance on N\'{e}el temperature, not only the large MAE but also the small R2 score. The reason for the failure may lie in the fact that the lengths of SM and ESM are too long on a small dataset, whose information is not relevant enough to the N\'{e}el temperature. The prediction error of OFM is lower than that of SM and ESM, which may be attributed to the orbital interactions encoded in OFM. The vector representation of OFM is also more compact than SM and ESM.

\begin{figure}[htbp]
	\centering
	\includegraphics[height=4.0cm,width=8.3cm]{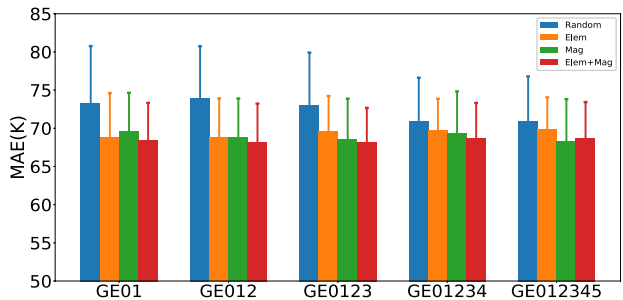}
	\caption{Prediction performance of the KRR model on the N\'{e}el temperature sub-dataset when Random-GEs, Elem-GEs, Mag-GEs, and Elem+Mag-GEs are taken as inputs.}
	\label{multi}
\end{figure}

\begin{figure}[htbp]
	\centering
	\includegraphics[height=5.0cm,width=8.0cm]{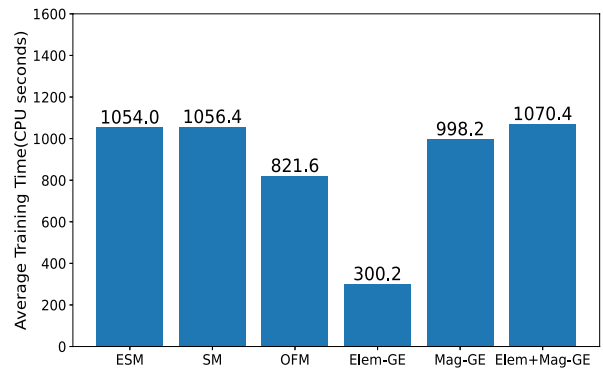}
	\caption{Average training time of the KRR model on five training sets, given different material representations as inputs.}
	\label{cost}
\end{figure}

Next, we examine the prediction performance of the KRR model on the N\'{e}el temperature sub-dataset, taking the Elem-GEs and Mag-GEs as inputs. In order to further verify the effects of self-supervised learning, we denote the multi-scale material descriptors generated by the randomly initialized GNN as Random-GE. We have a few remarks on the effects of self-supervised learning shown in Fig.\ref{multi}. (1) Random-GEs already have relatively good performance despite the absence of a specific type of knowledge through self-supervised training. The MAE obtained by Random-GE01 is 73.23 K, which is slightly better than OFM. However, the length of Random-GE01 is only 128, which is much shorter than 1,056, the length of OFM. The shorter length of the former indicates that Random-GE01 contains rich physical information more compactly. The reason is evident since the original crystal graph already contains sufficient information about materials, and atomic and material representations obtained by message passing on the crystal graph can naturally acquire information about elements and structures. (2) For material descriptors at each scale, the prediction error of self-supervised representations is systematically lower than that of random representations, i.e., the MAE of Elem-GEs or Mag-GEs is smaller than the MAE of Random-GEs, which proves that self-supervised pre-training is indeed beneficial. Specifically, we regularize the random weights in a more significant way for material properties by performing self-supervised training. Then we transfer it to the vector representation of the material and gain a more accurate machine learning model. (3) As we can see, by virtue of the combination of two types of knowledge, i.e., chemical rules and magnetism, combining the Elem-GEs and Mag-GEs (Elem+Mag-GEs) achieves the lowest prediction error on all scales except GE012345. The exception of GE012345 is due to the competition between the physical information in the descriptor and the vector length of the descriptor. (4) The self-supervised representations, Elem-GE012, Mag-GE012345, and Elem+Mag-GE0123, achieve the lowest prediction errors of 68.78 K, 68.34 K, 68.10 K, respectively. Their R2 scores are 0.32, 0.34, 0.32, respectively. The performance of the self-supervised representations is much better than popular manually constructed descriptors, e.g., ESM, SM, and OFM, at a  similar overall training cost of time, as shown in Fig.\ref{cost}.

\begin{table*}[htbp]
	\caption{Prediction performance of CGCNN and the KRR model on the full  N\'{e}el temperature dataset. OFM, Mag-GE01, Elem-GE01, and Elem+Mag-GE01 are taken as inputs of the KRR model.}
	\centering
	\def\temptablewidth{0.6\textwidth}
	{\rule{\temptablewidth}{1.0pt}}  
	\begin{tabular*}{\temptablewidth}{@{\extracolsep{\fill}}ccccccc}
		\hline
		&&OFM&CGCNN&Mag-GE01&Elem-GE01&Elem+Mag-GE01\\
		\hline
		&MAE&65.14$\pm$5.95&63.85$\pm$6.85&59.72$\pm$6.15&58.32$\pm$4.70&58.23$\pm$5.08\\
		\hline
		&R2&0.42$\pm$0.06&0.42$\pm$0.06&0.50$\pm$0.04&0.54$\pm$0.05&0.54$\pm$0.05\\
		\hline
	\end{tabular*}
	{\rule{\temptablewidth}{0.6pt}}
	\label{whole}
\end{table*}

After demonstrating the advantages of self-supervised multi-scale material representations over traditional descriptors SM and ESM on a sub-dataset of size 748, we train a better model by utilizing the whole N\'{e}el temperature dataset of size 1007 and compare it with OFM and CGCNN. As shown in Table \ref{whole}, the performance of OFM is the worst, which is consistent with our previous observations on the subset, illustrating the limitation of manually constructed descriptors. The powerful CGCNN also fails to achieve good prediction performance due to the limitation of the size of the dataset, and the sophisticated tuning procedure of hyperparameters makes CGCNN less efficient compared with the KRR model. However, taking the self-supervised multi-scale material representations as inputs of the KRR model, the prediction accuracy is better than CGCNN. The training is also more efficient and flexible.

\begin{figure}[htbp]
	\centering
	\includegraphics[height=6.0cm,width=8.0cm]{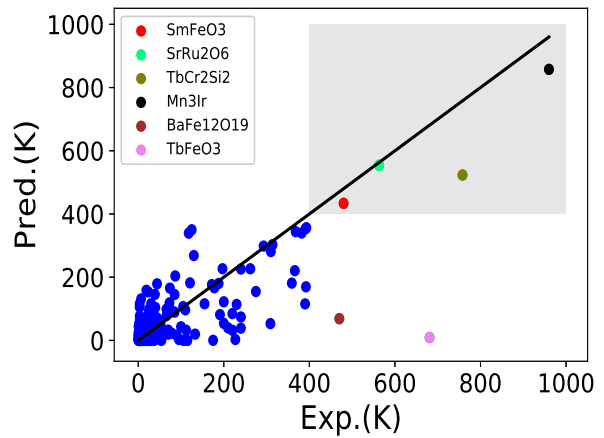}
	\caption{Parity plot of the prediction performance of the trained KRR model on the test set when Elem-GE01 are taken as inputs. The six antiferromagnetic materials in the high N\'{e}el temperature regime(Exp. $>$ 400 K) are colored differently, and the four successfully predicted materials are shown in the gray zone.}
	\label{parity}
\end{figure}

We can further analyze the trained model by drawing its parity plot, i.e., experimental values versus predicted values on the test set. For instance, the parity plot of Elem-GE01 on one of five test sets is shown in Fig.\ref{parity}. We focus on the predictions of antiferromagnetic materials with the experimental N\'{e}el temperatures in the high-temperature regime, i.e., above 400 K, since it is more relevant for applications in the field of antiferromagnetic spintronics\cite{JenkinsS}. We find that the predicted high N\'{e}el temperatures of most materials agree with the experimental high N\'{e}el temperatures, which is shown in the grey zone of Fig.\ref{parity}. For example, the predicted N\'{e}el temperatures are 433.65 K for SmFeO$_3$ and 553.99 K for SrRu$_2$O$_6$, respectively, with relative errors of only 9.6 \% to the experimental value 480 K of SmFeO$_3$\cite{KuoC} and 1.6\% to 563 K of SrRu$_2$O$_6$\cite{HileyC}. Particularly, the antiferromagnetic material Mn$_3$Ir\cite{TomenoI} has the highest N\'{e}el temperature of 960 K in the test set, and the trained KRR model also finds the highest predicted value of 857 K, with a relative error of only 10.6\%, showing that the trained model owns the ability to screen high-temperature antiferromagnetic materials in a high-throughput way. However, for BaFe$_{12}$O$_{19}$\cite{CollombA} and TbFeO$_3$\cite{BertautEF}, the predictions are completely failed. The dataset lacks sufficient materials covering more diverse chemical compositions and crystal structures. If we can expand the dataset by including more types of antiferromagnetic materials, exceptional failures like BaFe$_{12}$O$_{19}$ and TbFeO$_3$ would be erased.

\emph{Conclusion and outlook.---} We propose a self-supervised training strategy of CGCNN to extract material representations with rich physical information. The combination of the self-supervised material representations and the KRR model outperforms popular manually constructed descriptors as well as CGCNN on the N\'{e}el temperature dataset. The self-supervised GNN may also serve as a universal pre-training framework for various material properties. Although the trained model shows the ability to screen high N\'{e}el temperature materials, there are still challenges to gain a more reliable and accurate model. First, the proper encoding of the information of magnetic structures, i.e., the value and direction of magnetic moments, is of great significance since the magnetic materials with different N\'{e}el temperatures could have the same chemical composition and crystal structures but different magnetic structures. The self-supervised material descriptors cannot distinguish them, and geometric deep learning may be a promising solution\cite{BronsteinM}. Second, considering the absence of a large-scale, high-quality computational dataset of magnetic materials, we can further perform transfer learning\cite{BapstV,ZuoY,LeeJ} on low-fidelity datasets, which is also an effective strategy to improve the prediction performance of the model on small datasets.

\begin{acknowledgments}
	This work was supported by the Scientific Research Program from Science and Technology Bureau of Chongqing City (Grant No. cstc2020jcyj-msxmX0684), the Science and Technology Research Program of Chongqing Municipal Education Commission (Grant No. KJQN202000639), and in part by the National Natural Science Foundation of China under Grant No. 12147102.
\end{acknowledgments}

\newpage

\end{document}